\documentclass[sigconf,anonymous=False,nonacm]{acmart}

\AtBeginDocument{%
	\providecommand\BibTeX{{%
			\normalfont B\kern-0.5em{\scshape i\kern-0.25em b}\kern-0.8em\TeX}}}

\usepackage{comment}
\usepackage{multirow}
\usepackage{amsmath}
\usepackage{amsthm}
\usepackage{mathrsfs}
\usepackage{rotating}
\usepackage{xspace}
\usepackage{subcaption}
\usepackage[section]{placeins}
\def\Sys{URLTran\xspace}
\def\Sysb{URLTran\_BERT\xspace}
\def\Sysr{URLTran\_RoBERTa\xspace}
\def\Sysc{URLTran\_CustVoc\xspace}
\graphicspath{{figures/}} %Setting the graphicspath
\setcopyright{none}

\begin{document}
\title{\Sys : Improving Phishing URL Detection Using Transformers}

\author{Pranav Maneriker}
\authornote{Work done while the author was an intern at Microsoft Research.}
\authornotemark[2]
\email{maneriker.1@osu.edu}
\affiliation{The Ohio State University}
\author{Jack W. Stokes}
\email{jstokes@microsoft.com}
\affiliation{Microsoft}
\authornote{Both authors contributed equally to this research.}
\author{Edir Garcia Lazo}
\email{edirga@microsoft.com}
\affiliation{Microsoft}
\author{Diana Carutasu}
\email{dicaruta@microsoft.com}
\affiliation{Microsoft}
\author{Farid Tajaddodianfar}
\authornote{Work done while the author was employed at Microsoft.}
\email{f.tajad@gmail.com}
\affiliation{Amazon}
\author{Arun Gururajan}
\email{argurura@microsoft.com}
\affiliation{Microsoft}

\renewcommand{\shortauthors}{Maneriker et al.}
\begin{abstract}
Browsers often include security features to detect phishing web pages. In the past, some browsers evaluated an unknown URL for inclusion in a list of known phishing pages.
However, as the number of URLs and known phishing pages continued to increase at a rapid pace, browsers started to include one or more machine learning classifiers as part of their security services that aim to better protect end users from harm.
While additional information could be used, browsers typically evaluate every unknown URL using some classifier in order to quickly detect these phishing pages.
Early phishing detection used standard machine learning classifiers, but recent research has instead proposed the use of deep learning models for the phishing URL detection task.
Concurrently, text embedding research using transformers has led to state-of-the-art results in many natural language processing tasks.
In this work, we perform a comprehensive analysis of transformer models on the phishing URL detection task.
We consider standard masked language model and additional domain-specific pre-training tasks, and compare these models to fine-tuned BERT and RoBERTa models.
Combining the insights from these experiments, we propose \Sys which uses transformers to significantly improve the performance of phishing URL detection over a wide range of very low false positive rates (FPRs) compared to other deep learning-based methods.
For example, \Sys yields a true positive rate (TPR) of 86.80\% compared to 71.20\% for the next best baseline at an FPR of 0.01\%, resulting in a relative improvement of over 21.9\%.
Further, we consider some classical adversarial black-box phishing attacks such as those based on homoglyphs and compound word splits to improve the robustness of \Sys.
We consider additional fine tuning with these adversarial samples and demonstrate that \Sys can maintain low FPRs under these scenarios.

\end{abstract}

\maketitle

\section{Introduction}
\label{sec:intro}
Phishing occurs when a malicious web page is created to mimic the legitimate login page used to access a popular online service for the purpose of harvesting the user's credentials or a web page whose purpose is to input credit card or other payment information.
Typical phishing targets include online banking services, web-based email portals, and social media web sites.
Attackers use several different methods to direct the victim to the phishing site in order to launch the attack.
In some cases, they may send the user a phishing email containing the URL (Uniform Resource Locator) of a phishing page.
Attackers may also use search engine optimization techniques to rank phishing pages high in a search result query.
Modern email platforms use various machine learning models to detect phishing web page attacks.
In this work, we propose a new deep learning model that analyzes URLs and is based on transformers which have shown state-of-the-art performance in many important natural language processing tasks.

In order to prevent users from inadvertently uploading personal information to the attackers, web browsers provide additional security services to identify and block or warn a user from visiting a known phishing page.
For example, Google's Chrome browser utilizes their Safe Browsing technology~\cite{safebrowsing} and Microsoft's Edge browser includes Windows Defender SmartScreen~\cite{SmartScreen}.
In a related attack which is also addressed by these services, malicious URLs may point to a web page hosted by a misconfigured or unpatched server with the goal of exploiting browser vulnerabilities in order to infect the user's computer with malware (i.e., malicious software).

Successful phishing web page detection includes a number of significant challenges.
First, there is a huge class imbalance associated with this problem.
The number of phishing pages on the internet is very small compared to the total number of web pages available to users. Second, phishing campaigns are often short-lived. In order to avoid detection, attackers may move the login page from one site to another multiple times per day.
Third, phishing attacks continue to be a persistent problem.
 The number of known phishing sites continues to increase over time. Therefore, blocking phishing attacks only using a continuously growing list of known phishing sites often fails to protect users in practice.

Popular web browsers may render hundreds of millions or even billions of web pages each day.
In order to be effective, any phishing or malicious web page detection must be fast.
For this reason, several researchers~\cite{Blum2020,Le2018,Tajaddodianfar2020} have proposed detecting both phishing and malcious web pages based solely on analyzing the URL itself.
\begin{comment}
Phishing cyberattacks have been occurring for approximately 25 years, and their main goal remains the same today: convincing the user to disclose their credentials and financial information which the attacker can then use for their own financial gain. Phishing attacks can be initiated through emails and social media, and they target financial, online payments, social media, entertainment and technological companies. The attacker usually impersonates a known and reputable brand (e.g. Bank of America, Google, Amazon, Facebook, Microsoft) and attempts to convince the victim to disclose personal information.
\end{comment}

With the proliferation and ease of access to phishing kits sold on the black market as well as the phishing as a service offerings, it has become easy for attackers with little expertise to deploy phishing sites and initiate such attacks. Consequently, phishing is currently on the rise and costing over \$57 million from more than 114,000 victims in the US last year according to a recent FBI report~\cite{fbi2019}.
The number of phishing attacks rose in Q3 of 2019 to a high level not seen since late 2016~\cite{HelpNetSecurityPhish}.
As phishing is proving to be more and more fruitful, the attacks have become increasingly sophisticated. At the same time, the 
lifespan of phishing URLs has continued to drop dramatically – from 10+ hours to minutes~\cite{zvelophish}.

Given the significant repercussions of visting a phishing or malicious web page, the detection of these URLs has been an active area of research~\cite{Sahoo2017}.
In some cases, researchers have proposed the use of classic natural language processing methods to detect malicious URLs~\cite{Blum2020}.  Recently, others have begun to use deep learning models to detect these URLs. URLNet~\cite{Le2018} is a deep convolutional neural network (CNN) and includes separate character and word-level models for the malicious URL detection task.
The Texception~\cite{Tajaddodianfar2020} model, which is used to detect phishing URLs,
extends some of the ideas in URLNet by including small kernels which can be deployed in a wide variety of configurations in terms of width, depth or both.

Recently, semi-supervised machine learning methods have been used to create text embeddings that offer state-of-the-art results in many natural language processing tasks.
The key idea in these approaches is the inclusion of a transformer model~\cite{vaswani2017attention}. BERT~\cite{devlin2019bert,rogers2020primer} utilizes transformers to offer significant
improvements in several natural language processing (NLP) tasks. GPT~\cite{alec2018improving}, GPT-2~\cite{gpt2}, and GPT-3~\cite{gpt3} have also followed a similar approach.
The semantics and syntax of natural language are more complex than URLs, which must follow a strict syntax specification~\cite{rfc3986}.
However, recent work using transformers has also demonstrated that these models can be applied to tasks involving data with more strict syntactic structures.
These include tabular data~\cite{yin2020tabert}, python source code~\cite{kanade2020pre} and SQL queries~\cite{wang2020rat}.
The success of these approaches further motivates us to apply transformers on URLs.

In this paper, we compare two settings: 1) we pre-train and fine-tune an existing transformer architecture using only URL data, and 2)
we fine-tune publicly available pre-trained transformer models.
In the first approach, we apply the commonly used Cloze-style masked language modeling objective~\cite{taylor1953cloze} on  the BERT architecture. 
In the second approach, we fine-tune BERT~\cite{devlin2019bert} and RoBERTa~\cite{liu2019roberta} on the URL classification task.
Each of these systems forms an example of a \Sys model.
\mbox{\Sysb} is the best performing model obtained from these approaches.
Finally, we simulate two common black-box phishing attacks by perturbing URLs in our data using unicode-based homoglyph substitutions~\cite{woodbridge2018detecting} and inserting `-' characters between sub-words in a compound URL (e.g., `bankofamerica.com' $\rightarrow$ `bank-of-america.com'), along with a perturbation scenario under which the parameters are reordered and the URL label remains unchanged to improve the robustness of \Sys.

Results on a large corpus of phishing and benign URLs show that transformers
are able to significantly outperform recent state-of-the-art phishing URL detection models (URLNet, Texception)
over a wide range of low false positive rates where such a phishing URL detector
must operate. At a false positive rate of 0.01\%,
\Sys increases the true positive rate from 71.20\% for the next best baseline (URLNet)
to 86.80\% (21.9\% relative increase). Thus, browser safety services, such Google's Safe Browsing and Microsoft's SmartScreen, may potentially benefit using the proposed \Sys system for the detection of phishing web pages.

This paper offers the following contributions:
\begin{itemize}
	\item Borrowing from recent advances in many natural language processing tasks, we
	propose the use of transformers to improve the detection of phishing URLs.
	\item We build \Sys, a large-scale system with production data and labels and demonstrate
	that transformers do offer a significant performance improvement compared to previous recent deep learning solutions over a wide range of very low false positive rates.
	\item We analyze the impact of various design choices in terms of hyperparameters, pre-training tasks, and tokenizers to contribute to an improved model.
	\item We analyze the adversarially generated URLs  from the system to understand
	the limitations of \Sys. 
\end{itemize}

\section{Phishing URL Data}
\label{sec:data}
The datasets used for training, validation and testing were collected from
Microsoft’s
Edge and Internet Explorer
production browsing telemetry during the summer of 2019. The schema for all three datasets is similar and consists of the browsing URL and a boolean determination of whether the URL has been identified as phishing or benign.

Six weeks of historical data were collected overall out of which four weeks of data were used for the training set, one week for the validation and one week for the test set.

Due to the highly unbalanced nature of the datasets (roughly 1 in 50 thousand URLs is a phishing URL), down-sampling of the benign set was necessary and resulted in a ratio of 1:20 (phishing versus benign) for both the training and validation sets. The resulting datasets had the corresponding total sizes of 1,039,413 records for training and 259,854 thousand for validation.
The test set used for evaluating the models consists of 1,784,155 records, of which 8,742 are phishing URLs and the remaining 1,775,413 are benign.

The labels included in this study correspond to those used to train production classifiers.
Phishing URLs are manually confirmed by analysts including those which have been reported as suspicious by end user feedback. Other manually confirmed URLs are also labeled as phishing when they are included and manually verified in known phishing URL lists including Phishtank. \footnote{The total of $73,705$ valid phishing URLs is significantly larger than the number of phishing URLs reported by Phishtank (\url{http://phishtank.org/stats.php}).}

Benign URLs are those which correspond to web pages which are known to not be involved with a phishing attack. In this case, these sites have been manually verified by analysts using manual analysis.
In other cases, benign URLs can be confirmed by thorough (i.e., production grade) off-line automated analysis which is not an option for real-time detection required by the browser.
None of the benign URLs have been included in known phishing lists or have been reported as phishing pages by users and later verified by analysts. Although these last two criteria are not sufficient to add an unknown URL to the benign list.
It is important to note that all URLs labeled as benign correspond to web pages that have been validated.
They are not simply a collection of unknown URLs, i.e., ones which have not been previously detected as phishing sites. 

\section{Methodology}
\label{sec:method}
\Sys seeks to use recent advances in natural language processing to improve the task of detecting phishing URLs.
Building \Sys employs a two-pronged approach towards adapting transformers for the task of phishing URL detection.
First, state-of-the-art transformer models, BERT~\cite{devlin2019bert} and  RoBERTa~\cite{liu2019roberta}, are fine-tuned, starting from publicly available vocabularies and weights and across different hyperparameter settings and resulting in \Sysb and \Sysr, respectively.
Second, domain-specific vocabularies are built using different tokenization approaches, and a domain specific transformer (\Sysc) is first pre-trained and then fine-tuned on the task. 

\begin{figure}
	\centering
	\includegraphics[width=0.5\linewidth]{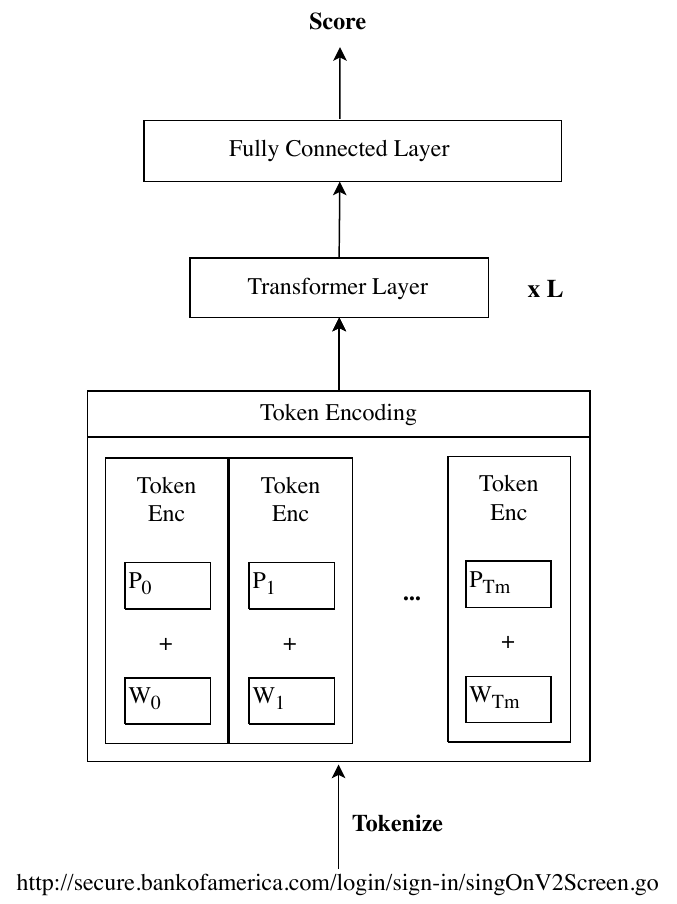}
	\caption{\Sys phishing URL detection model.}
	\label{fig:transformer}
\end{figure}

 The general architecture of all the explored models takes a three stage approach for inference shown in Figure~\ref{fig:transformer}. It first uses a subword tokenizer to extract tokens from a URL.
Next, a  transformer model generates an embedding vector for the unknown URL.
Finally, a classifier predicts a score indicating whether or not the unknown URL corresponds to a phishing web page.

In the following sections, we first provide briefly summarize the transformer model architecture, followed by the training tasks used to train the model, and end with a description of the adversarial settings under which the best \Sys model is evaluated and then trained with adversarial examples to improve its robustness.

\subsection{Architecture}
We describe the tokenization schemes and overall architecture for classification in this section, skipping a detailed description of transformer models for brevity.
Interested readers can review the transformer~\cite{vaswani2017attention}, BERT~\cite{devlin2019bert}, or RoBERTa~\cite{liu2019roberta} papers for details of the internal structure of transformer layers.
\subsubsection{Tokenization}
The raw input to the \Sys model is the URL, which can be viewed as a text sequence.
The first step in the phishing URL detection task involves converting this input URL into a numerical vector which can be further processed by a classic machine learning or deep learning model.

Previous URL detection models~\cite{Blum2020} extracted lexical features by first splitting the URL with a set
of important delimiters (e.g., `=', `/', `?', `.', ` ') and then creating a sparse binary features based on these tokens.
Recent deep learning-based URL detection models~\cite{Le2018,Tajaddodianfar2020} instead include separate word-level and character-level CNNs where the character-level CNNs span different lengths of character subsequences.

Instead of these approaches, we experiment with multiple subword tokenization schemes in \Sys.
Subword models have seen increased adoption in different tasks in NLP, including machine translation~\cite{sennrich2016neural}, word analogy~\cite{bojanowski2017enriching}, and question answering~\cite{zhang2019effective}.
While using full-length words reduces the input representation length (number of tokens) allowing more input to be processed by a fixed-length model, using a subword model can provide morphological insights to improve inference.
For example, a full-length model would consider `bankofamerica' and `bankofcanada' as completely unrelated tokens, whereas a subword model can recognize the shared subword `bank' to correlate URLs belonging to the two banks.
Important character subsequences, including prefixes and suffixes can also provide relevant information while being more robust to polymorphic attacks.

In particular, for \Sysb and \Sysr, we use the existing word piece~\cite{Wu2016GooglesNM,devlin2019bert}  and Byte Pair Encoding (BPE) models~\cite{gpt2,liu2019roberta} , respectively.
In addition to these, custom character-level and byte-level BPE vocabularies are created using the training URL data to have a domain specific vocabulary for \Sysc with two different vocabulary sizes,  1K and 10K.
The BPE models attempt to find a balance of using both character subsequences and full words.

The BPE models first break the $m^{th}$ URL, $u_m$, into a sequence of text tokens, $\mathbf{TOK}_m$, where
the individual tokens  may represent entire words or subwords~\cite{Schuster2012,Sennrich2016,Wu2016GooglesNM}.
Following the notation in~\cite{devlin2019bert}, the token sequence is formed as:
\begin{equation}
	\begin{split}
		& \mathbf{TOK}_m = \text{Tokenizer}(u_m)\\
	\end{split}
\label{eq:tok}
\end{equation}
%\end{comment}
\noindent where $\mathbf{{TOK}_m}$ is of length $T_m$ positions and consists of individual tokens ${Tok}_t$ at each position index $t$.
%${Tok}_t$ is an individual token generated from the $m^{th}$ URL generated by the wordpiece model.
For example, the BERT wordpiece token sequence generated from the URL of a popular banking login page,\\
\mbox{\small{$u_m$ = secure.bankofamerica.com/login/sign-in/signOnV2Screen.go}} \\
 is shown in Table~\ref{tab:boa}. The wordpiece model includes special text tokens specified by (\#\#) which build
 upon the previous token in the sequence. In the example in Table~\ref{tab:boa}, `\#\#of' means that it occurs after a previous
 token (`bank'), and it is distinguished from the more common, separate token `of'.

\begin{table*}
\centering
\begin{tabular}{|l|l|}
\hline
URL ($u_m$) & secure.bankofamerica.com/login/sign-in/signOnV2Screen.go \\
\hline
Token Sequence ($\mathbf{{TOK}_m}$) & \{ `secure', `.', `bank', `\#\#of', `\#\#ame', `\#\#rica', `.', `com', `/', `log', `\#\#in', `/', \\  & `sign', `-', `in', `/', `sign', `\#\#on', `\#\#v', `\#\#2', `\#\#screen', `.', `go' \} \\
  \hline
\end{tabular}
\caption{Example of the wordpiece token sequence extraction from a popular banking web page.}
\label{tab:boa}
\end{table*}

\subsubsection{Classifier}
The final encoding produced by the transformer model can be used for a variety of downstream NLP tasks such
as language understanding, language inference, and question answering, and text classification.
We use the transformer embeddings for two tasks: pre-training masked language models, and fine-tuning for classification of phishing URLs.
Both of these tasks require a final classification layer,  which can be applied to multiple tokens for masked token prediction, and a pooled representation for classification.
The transformer models that we train use a single, dense two-class classification layer, which is applied to a special pooled token (`\texttt{[CLS]}') for classification. A dense layer having \texttt{vocab\_size} classes is used for predicting the masked token for the masked language modeling task during pre-training:

\begin{equation}
	\begin{split}
		& \mathbf s_m = \mathbf W \mathbf x_m + \mathbf b. 
	\end{split}
    \label{eq:dense}
\end{equation}

\noindent  In ~(\ref{eq:dense}), $\mathbf{W}$ and $\mathbf{b}$ are the weight matrix and bias vector, respectively, for the final
dense linear layer.
 $\mathbf{s_m}$ is the
score which predicts if the URL $\mathbf{u_m}$ corresponds to a phishing web page when performing classification and is the sequence of masked token probability score vectors when performing masked language modeling for input token $\mathbf{x_m}$.

\subsection{Training}
\subsubsection{Masked Language Modeling (MLM)}
The MLM task is commonly used to perform pre-training for transformers. In this task, a random subset of tokens is replaced by a special `\texttt{[MASK]}' token.
 The training objective for the task is the cross-entropy loss corresponding to predicting the correct tokens at masked positions.
 The intuition for using this task for URLs is that specific query parameters and paths are generally associated with non-phishing URLs and therefore predicting masked tokens would help to uncover these associations.
 Similar intuitions derived from the cloze task~\cite{taylor1953cloze} motivate the usage of MLMs for pre-training natural language models.
  Following the MLM hyperparameter settings for BERT, 15\% of the tokens were uniformly selected for masking, of which 80\% are replaced, 10\% were left unchanged, and 10\% were replaced by a random vocabulary token at each iteration.
  Dynamic masking~\cite{liu2019roberta} was used, i.e., different tokens masked from the same sequence across iterations.
 The training subset of the full dataset was used for pre-training to prevent any data leakage. 

\subsubsection{Fine Tuning}
For \Sysb and \Sysr, all of the initial parameters derived using a large, internal natural language corpus generated by their respective authors, were used.
For \Sysc, following the completion of the MLM pre-training step, the learned weights were used as initialization values.

Next, \Sys's model parameters were further improved
using a second ``fine-tuning'' training process which depends upon the error
signal from the URL classification task and gradients based on gradient descent using the Adam~\cite{kingma2014adam} optimizer with the cross-entropy loss.

\subsection{Adversarial Attacks and Data Augmentation}
\label{sec:adv_attack}

Phishing URL attacks can occur on short-lived domains and URLs which have small differences from existing, legitimate domains.
We simulate two attack scenarios by constructing examples of such adversaries based on modifying benign URLs.
Note that these generated domains do not actually exist in the pre-existing training and testing data, but are based upon frequently observed phishing attack patterns.
We also utilize a reordering-based augmentation, which is used is used to generate benign perturbations for evaluating adversarial attacks.

\subsubsection{Homoglyph Attack}
We generate domains that appear nearly identical to legitimate URLs by substituting characters with other unicode characters that are similar in appearance.
This attack strategy is commonly referred to as a \textit{homoglyph attack}~\cite{yerima2020high,JiDeepWordBug18}, and
we implement this strategy using the python library \texttt{homoglyphs}\footnote{https://pypi.org/project/homoglyphs/}.
In particular, given a URL, we first extract the domain.
For a randomly selected character in the domain, we check for one unicode (utf-8) Latin or Cyrillic character that is a homoglyph for it. We only perturb one character to minimize
the probability that such a URL would we be identified as phishing by the user.
We then replace the character by its homoglyph to construct a new URL.
The URLs generated from this strategy are labeled as phishing.

\subsubsection{Compound Attack}
An alternative way to construct new phishing URLs is by splitting domains into sub-words (restricted to English) and then concatenating the sub-words with an intermediate hyphen.
For example, `bankofamerica.com' $\rightarrow$ `bank-of-america.com'.
To implement this, we leverage the \texttt{enchant} dictionary\footnote{https://pypi.org/project/pyenchant/}.
Consider a URL with domain $d$ having $|d| = n$ characters.
Let $\mathscr{D}$ denote the \texttt{enchant} English dictionary.
Let $C(d, i, j)$ denote the function that returns True if $d[i \dots j]$ can be split into one or more parts, each of which is a word in the dictionary $\mathscr{D}$.
The compound word problem can be formulated recursively as
\begin{align}
	C(d, i, j) =
	\begin{cases}
		\text{True}, & d[i \dots j] \in \mathscr{D}\\
		\text{True} & \exists k, C(d, i, k) \text{ and } C(d, k+1, j)\\
		\text{False} & \text{otherwise}
	\end{cases}
\end{align}
Using this recursive definition, we implement a dynamic programming algorithm that can compute whether a domain can be split and the corresponding splits.
These splits are then concatenated with hyphens between the discovered words.
Note that the base case check  $d[i\dots j] \in \mathscr{D}$ is performed in a case insensitive manner to ensure that the dictionary checks do not miss proper nouns.

\newcommand{\permuteSys}{PermuteURL\xspace}
\subsubsection{Parameter Reordering}
\begin{figure}
	\centering
	\includegraphics[width=0.75\linewidth]{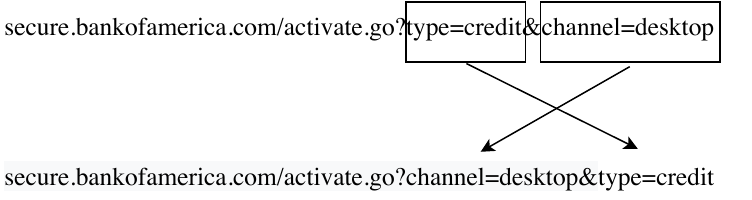}
	\caption{An example of parameter reordering}
	\label{fig:permute_url}
\end{figure}

Data augmentation using invariants, contextual replacement, and reward-based learning~\cite{kobayashi2018contextual,hu2019learning} has been used to improve classifiers in the text domain.
These can be extended to augment data in the URL domain.
As the query parameters of a URL are interpreted as a key-value dictionary, this augmentation incorporates permutation invariance. An example of a URL and permutation is provided in Figure~\ref{fig:permute_url}.
We use this approach to generate benign examples.
Reordering the parameters still results in a valid URL, i.e., parameter reordering does
not represent a phishing attack, and therefore we do not modify the URL's label.

\subsubsection{Adversarial Attack Data}
The approach we use for generating data for an adversarial attack includes generating separate augmented training, validation and test datasets based on their original dataset~\cite{Goodfellow2014}.
For each URL processed in these datasets, we generate
a random number. If it is less than 0.5, we augment the URL, or otherwise, we include it in its original form. For URLs which are to be augmented,
we modify it using either a homoglyph attack, a compound attack or parameter reordering with equal probability. If a URL has been augmented,
we also include the original URL in the augmented dataset.

\section{Numerical Evaluation}
\label{sec:eval}
In this section, we present the numerical evaluation of the different approaches presented in the previous sections.
We then  compare \Sys to several recently proposed baselines. We also report the the model's training and inference times.
Finally, we analyze the robustness of the model to  generated phishing URLs.

\noindent\textbf{Setup.}
The hyperparameter settings for all models are provided in Appendix~\ref{sec:hyper}.
In our experiments, we set the hyperparameters for previously published models according to their settings in the original paper.
For evaluating \Sysc, we vary the number of layers between $\{3, 6, 12\}$, vary the number of tokens per input URL sequence between $\{128, 256\}$, and  use both a byte-level and character-level BPE tokenizer with 1K- and 10K-sized vocabularies.
We randomly pick 15 hyperparameter combinations among these settings and present the results for these.
The Adam optimizer~\cite{kingma2015adam} is used in both pre-training and fine-tuning, with the triangular scheduler~\cite{smith2017cyclical} used for fine-tuning.

All training and inference experiments were conducted
using PyTorch~\cite{PyTorch} version 1.2 with NVIDIA Cuda 10.0 and Python 3.6.
The experiments were performed by extending the Hugging Face and Fairseq PyTorch implementations found on GitHub~\cite{HuggingFace,Fairseq}.
The large class imbalance makes accuracy a poor metric of model performance.
We evaluated all the models using the true positive rate (TPR) at low false positive rate (FPR) thresholds. We used the receiver operating characteristics (ROC) curve to compute this metric.

\noindent\textbf{Baselines.}

To evaluate the performance of our models, we compared them to two baseline URL detection models: URLNet
and Texception.
URLNet~\cite{Le2018}
is a CNN-based model which was recently proposed for the task of detecting URLs which identify
malicious web sites. In our baseline, we have completely trained and tested the URLNet model for the
detection of phishing URLs.
Texception~\cite{Tajaddodianfar2020} is another deep learning URL detection
model which has been proposed for the task of identifying phishing URLs. It is important to note
that Tajaddodianfar et. al.~\cite{Tajaddodianfar2020} compared Texception to a Logistic Regression-based
model and found that Texception offered better performance. Thus, we did not repeat that baseline experiment in
this work.

\begin{figure}
\begin{subfigure}{\linewidth}
	\centering
\includegraphics[width=0.75\linewidth]{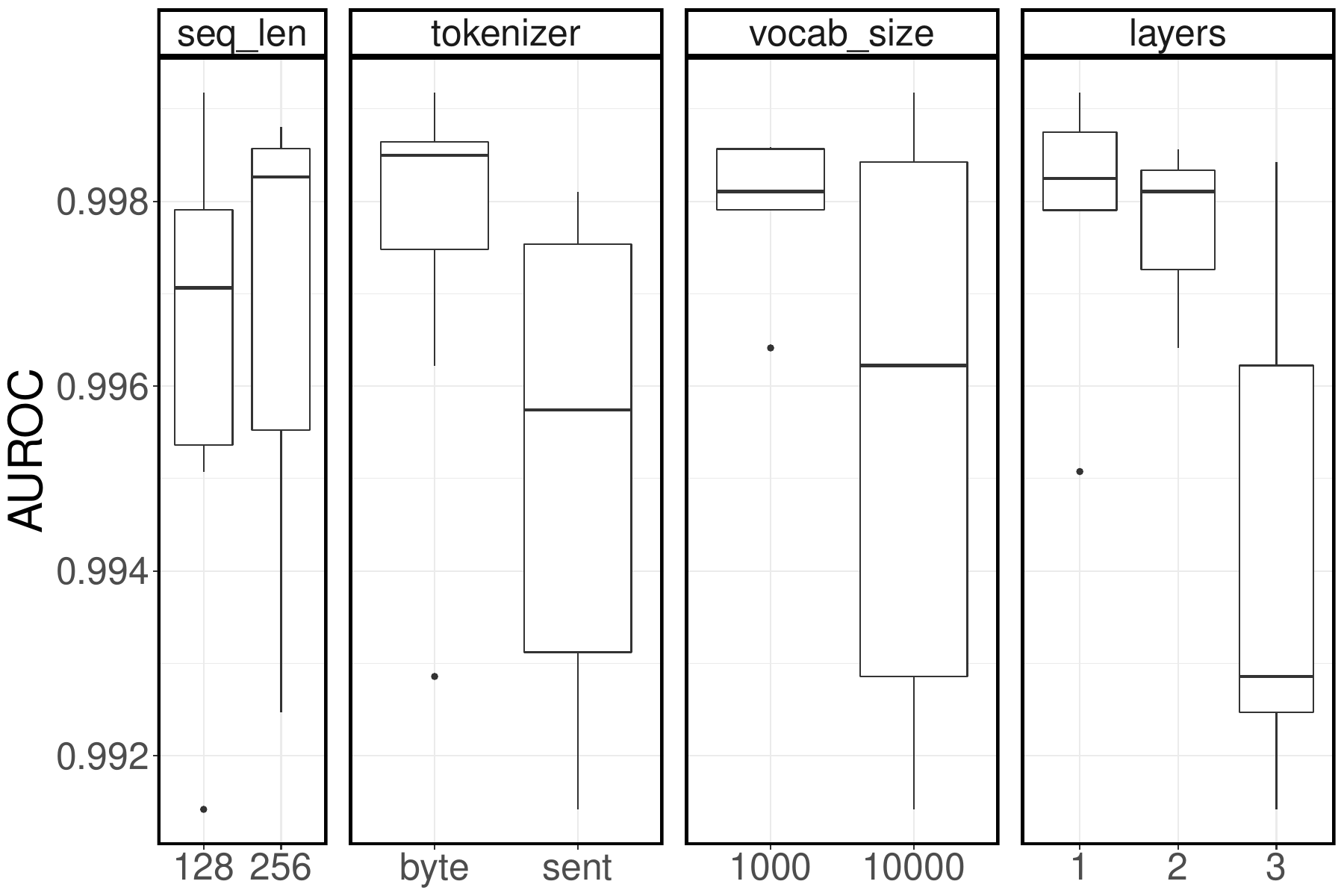}
\caption{Area under ROC vs hyperparameters}
\label{fig:pretrain_scratch:roc}
\end{subfigure}
\begin{subfigure}{\linewidth}
	\centering
\includegraphics[width=0.75\linewidth]{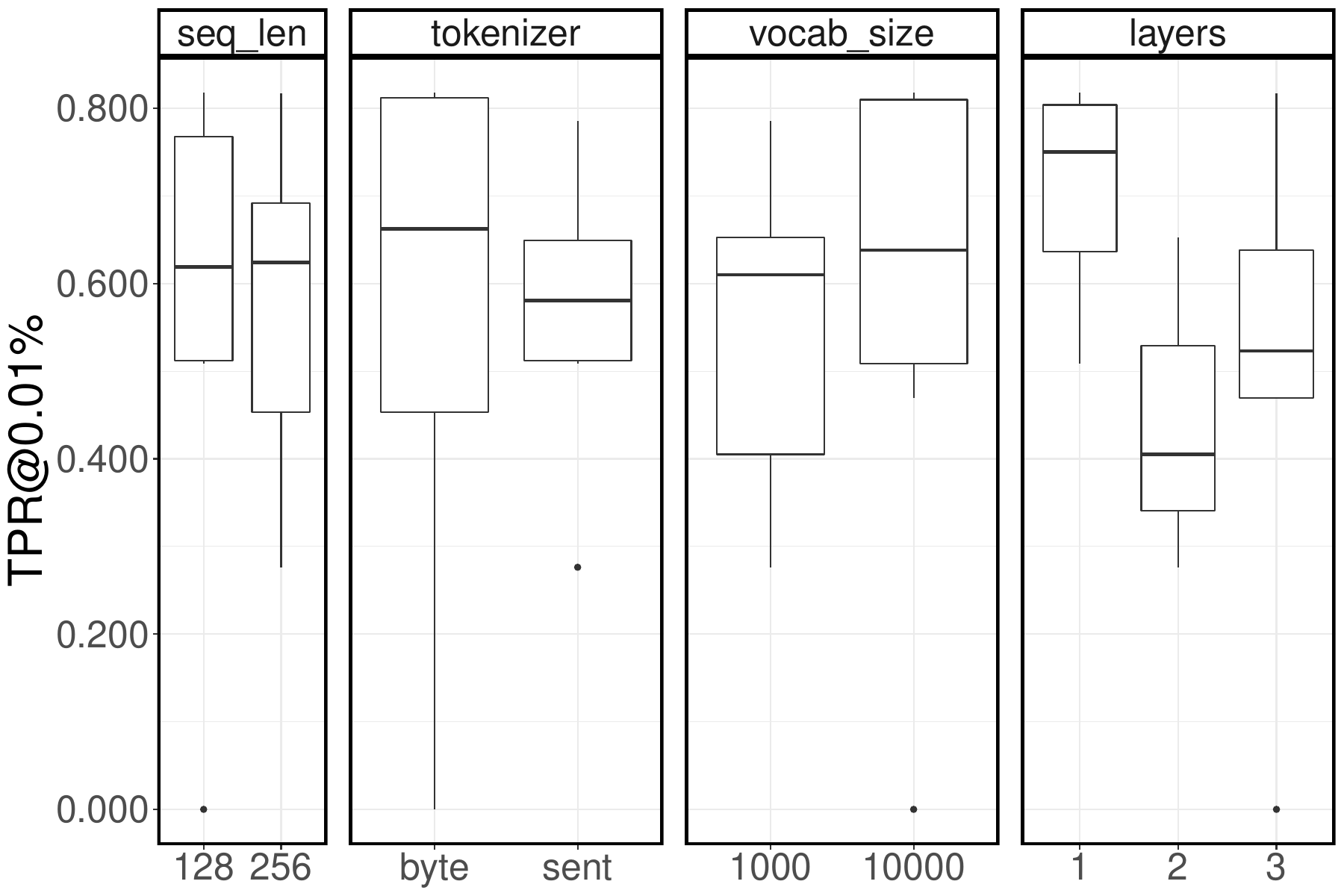}
\caption{TPR@FPR = 0.01\% vs hyperparmeters}
\label{fig:pretrain_scratch:tpr}
\end{subfigure}
\caption{Variance in quality of \Sysc across different hyperparameter settings}\
\label{fig:pretrain_scratch}
\end{figure}

\noindent\textbf{\Sysc.}
Transformers typically require large amounts of pre-training data (e.g., BERT~\cite{devlin2019bert} used a corpus of  $\approx$ 3.3 B tokens).
However, this data is derived from text articles, which are structured differently from URLs. We also trained the \Sysc model based soley on 
the URL data found in our datasets to compare the results of finetuning using standard BERT and RoBERTa pretrained models to models pretrained from the 
URL data. 
The difference in dataset size and data domain make it important to understand the impact of different hyperparameters used when training transformers from scratch.
We compare runs across different hyperparameters on the basis of area under ROC (AUROC) and TPR@0.01\% FPR.
Figure~\ref{fig:pretrain_scratch} demonstrates that the training is not very sensitive to sequence length. 
Smaller byte-level vocabularies tend to be better overall, but at low FPR, the difference is not significant.
Finally, we found that the 3 layer model generalized the best.
We hypothesize that the better performance of the model with fewer layers is because of limited pre-training data and epochs.
In the next few sections, we validate this hypothesis by evaluating fine-tune models that have have longer pre-training (\Sysb, \Sysr)  and that are tuned on a larger, adversarial dataset.

\begin{figure}
    \centering
	\includegraphics[width=0.9\linewidth]{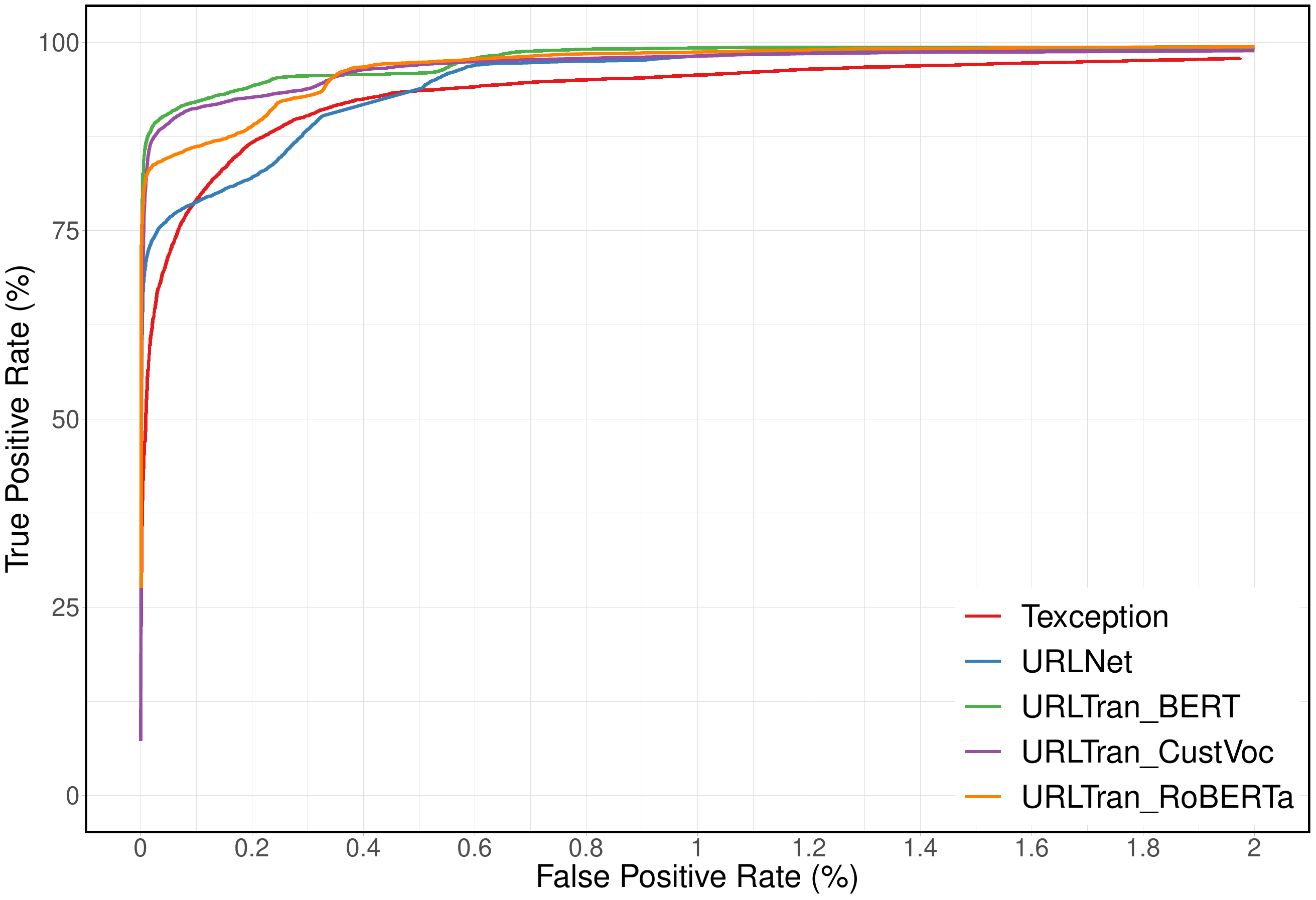}
	\caption{Receiver operating characteristic curve indicating the performance of the \Sys and several baseline models zoomed into a maximum of 2\% false positive rate.}
	\label{fig:transformer2}
\end{figure}

\begin{figure}
    \centering
	\includegraphics[width=0.9\linewidth]{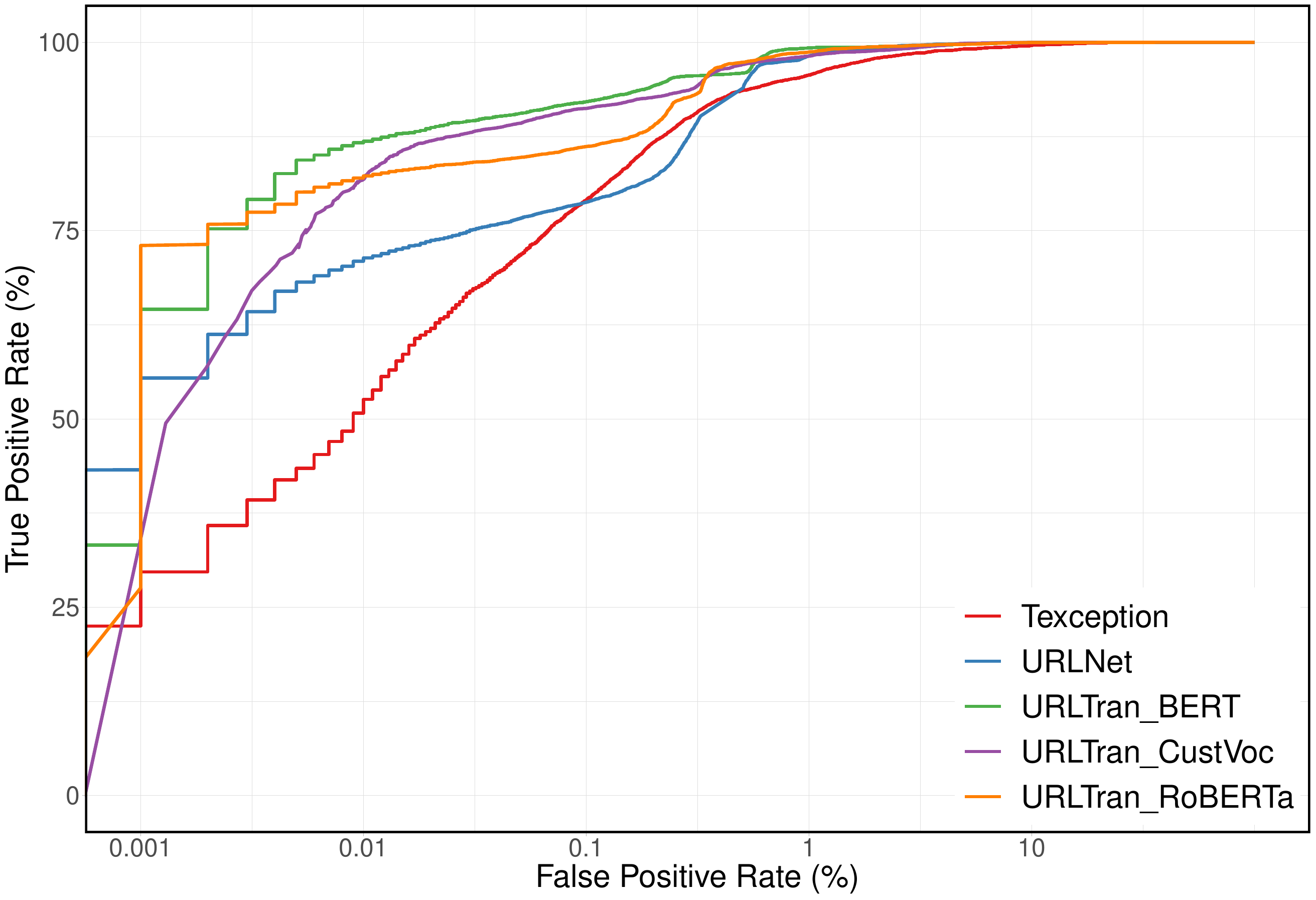}
	\caption{Zoomed in receiver operating characteristic curve with a log x-axis.}
	\label{fig:log_transformer2}
\end{figure}

\noindent\textbf{Model Performance.}
We next analyze the performance of the best parameters of all the proposed transformer variants.
To understand how these models compare
at \textit{very low FPRs} where detection thresholds must be set to operate in a production environment, we first
plot the ROC curves on a linear x-axis zoomed into a 2\% maximum FPR in Figure~\ref{fig:transformer2}.
We also re-plot these ROC curves on a log x-axis in the semilog plot in Figure~\ref{fig:log_transformer2}.
These results indicate that all variants of \Sys offer a significantly better
true positive rate over a wide range of extremely low FPRs. In particular, \Sys matches or exceeds the TPR of URLNet
for the FPR range of 0.001\% - 0.75\%. The result is very important because phishing URL detection models
must operate at very low FPRs (e.g., 0.01\%) in order to minimize the number of times the security service predicts that a benign
URL is a phishing site (i.e., a false positive). In practice, the browser manufacturer selects the desired FPR and tries
to develop new models which can increase the TPR for the selected FPR value.
Note that TPR@FPR is the \textbf{standard metric} commonly used both in production settings and in prior art such as Texception and URLNet.

In addition to the ROC curve analysis, we also summarize a number of key performance metrics in Table~\ref{tab:model-perf}.
In the table, `F1' is the F1 score, and `AUC' is the area under the model's ROC curve.
The proposed \Sys model outperforms both Texception and URLNet for all of these metrics.
In particular, we note that  at an FPR
of 0.01\%, \Sysb has a
TPR of 86.80\% compared to 71.20\% for URLNet and 52.15\% for Texception.

\begin{table*}[tb]
\begin{center}
\begin{tabular}{| c | c | c | c | c | c | c |}
\hline
Model & Accuracy (\%) & Precision (\%) & Recall (\%) & TPR@FPR=0.01\% & F1 & AUC \\
\hline
\hline
Texception & 99.6594 & 99.7562 & 99.6594 & 52.1505 & 0.9969 & 0.9977 \\
URLNet & 99.4512 & 99.7157 & 99.4512 & 71.1965 & 0.9954 & 0.9988 \\
URLTran\_CustVoc & 99.5983 & 99.7615 & 99.5983 & 81.8577 & 0.9965 & 0.9992 \\
URLTran\_RoBERTa & 99.6384 & 99.7688 & 99.6384 & 82.0636 & 0.9968 & 0.9992 \\
URLTran\_BERT & 99.6721 & 99.7845 & 99.6721 & 86.7994 & 0.9971 & 0.9993 \\
\hline
\end{tabular}
\end{center}
\caption {Comparison of different performance metrics for \Sys and the two baseline models}
\label{tab:model-perf}
\end{table*}

\noindent\textbf{Training and Inference Times.}
The total time required to train the best \Sysb model was 4:57:11 on an NVIDIA V100. Inference
required 0:10::44 to complete for an average of 0.36096 milliseconds per sample.

\begin{figure}
    \centering
	\includegraphics[width=0.9\linewidth]{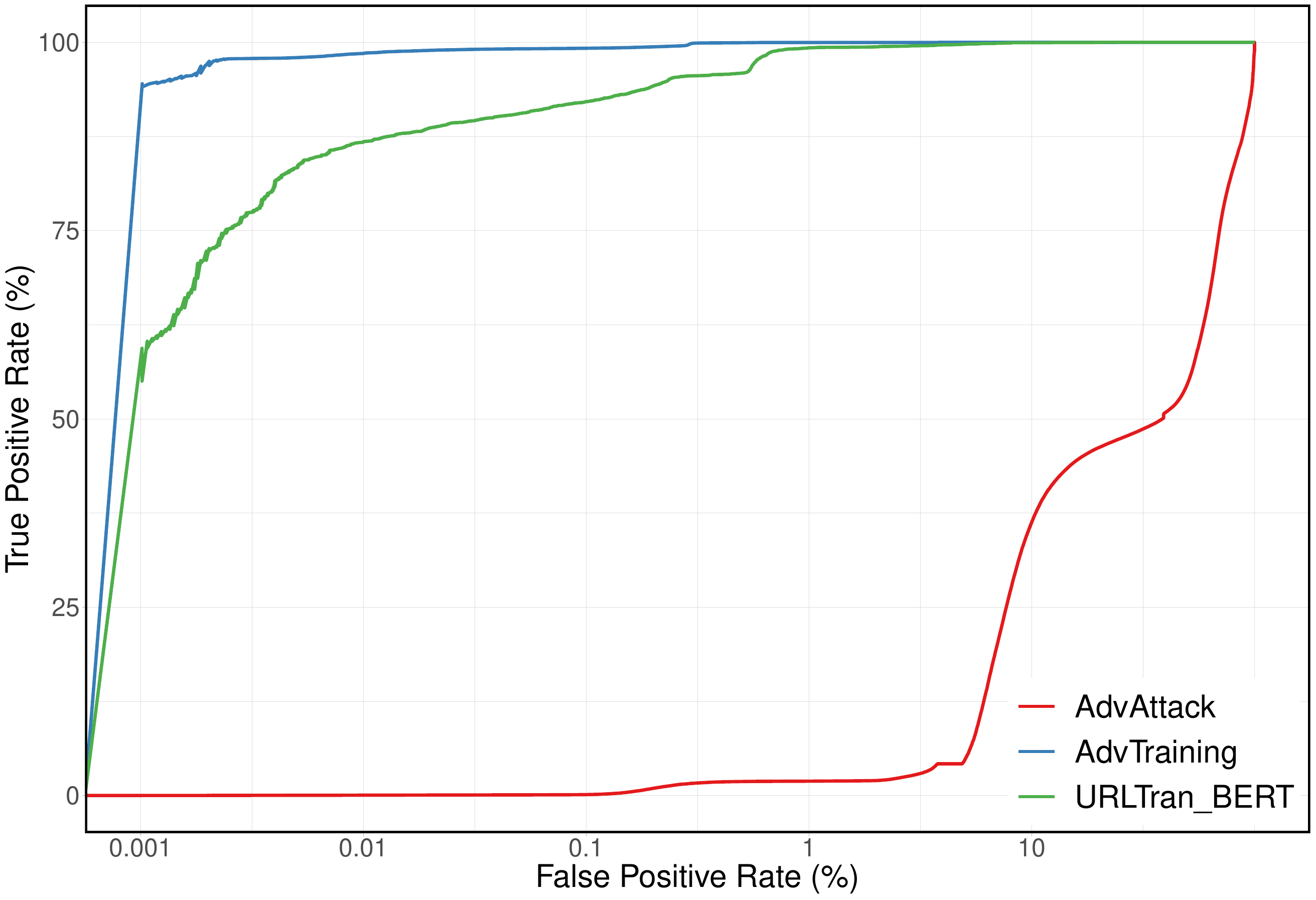}
	\caption{ROC curve for \Sysb when under adversarial attack, and adversarial robustness after augmented training}
	\label{fig:adversarial_roc}
\end{figure} 
\noindent\textbf{Adversarial Evaluation.}
To understand \Sys's robustness to adversarial attacks, we first compared the low FPR regions of the ROC curve of the unprotected model tested with the original test set to the
test set which includes adversarial samples (AdvAttack) generated through the methods described in Section~\ref{sec:adv_attack} (Figure~\ref{fig:adversarial_roc}). There is a significant drop in performance of \Sysb when attacked with adversarial URLs.
Next, we consider the scenario where  attack strategies are incorporated into the training data (AdvTraining).
On the addition of adversarial attack patterns to the training, the model is able to  the adapt to novel attacks, and even outperform the unprotected version of \Sys.
These results demonstrate that \Sys can adapt to novel attacks.
Further, as new attack strategies are recognized (e.g., homoglyph), a robust version of \Sys can be trained to recognize similar patterns in unseen test data.

\section{Related Work}
\label{sec:related}
The \Sys system is most closely related to phishing and malicious URL detection models which have been previously proposed in the
literature.
In this section, we describe related work for
deep learning-base text embeddings in general. These
models have been derived for various natural language processing tasks.
We then review related work in phishing and malicious web page detection using the
web page's URL which builds upon the previous text embedding models
proposed in the NLP domain. In particular, we focus on two recent, deep learning URL detection
models, URLNet and Texception, which helped to inspire this work.

\noindent\textbf{Text Embeddings.}
Deep learning models for text embeddings have been an active area of
research recently.
One form of models called a character-level CNN
learns a text embedding from individual characters, and these
embeddings are then processed using a sequential CNN and one or more dense
layers depending on the task. Recent examples
of character-level CNNs include~\cite{Conneau2017,Zhang2015}.
In particular, Conneau et al.~\cite{Conneau2017} investigated very deep architectures
for the purpose of classifying natural language text.
Typically, these models are trained in an
end-to-end fashion instead of from manually engineered features.

Transformers were introduced by Vaswani et al.~\cite{vaswani2017attention} in the context of neural
machine translation.
A number of models used transformers for other natural language processing
tasks including BERT~\cite{devlin2019bert,rogers2020primer}, GPT~\cite{alec2018improving}, GPT-2~\cite{gpt2}, and GPT-3~\cite{gpt3}.
RoBERTa~\cite{DBLP:journals/corr/abs-1907-11692} used careful optimization of the
BERT parameters and training methodology to offer further improvements.

\noindent \textbf{ Adversarial Attacks on Text.}
Adversarial example generation has been a focus of some recent work on understanding the robustness of various text classification tasks.
The examples generated using these approaches aim to impose certain semantic constraints without modifying the label of the underlying text.
White-box attacks (e.g., Hotflip~\cite{ebrahimi2018hotflip})  require access to the internals of the classification model used, such as the gradient on specific examples.
The attack framework proposed in our work is more in line with  black-box attack frameworks such as DeepWordBug~\cite{JiDeepWordBug18} and TextAttack~\cite{morris2020textattack}  where the construction of adversarial data is motivated by a threat model but independent of the classifier used.
We specialize this attack scheme for the URL context.

\noindent\textbf{URL-Based Phishing and Malicious Web Page Detection.}
Previous related work on the detection of phishing and malicious web pages
based on the page's URL has progressed in parallel. We next review some important
systems in chronological order.

Early phishing page detection based on URLs followed conventional deep learning
approaches. A summary of these methods is included in~\cite{Sahoo2017}.
Blum et al.~\cite{Blum2020} proposed using confidence weighted, online learning using a set of lexical features which
are extracted from the URL. To extract these features, the URL is first split using the following
delimiters:  `?', `=', `/', `.',  and ` '. Next, individual features are set based on the
path, domain, and protocol.

Le et al.~\cite{Le2018} proposed the URLNet model whose task is to detect URLs which are references
to malicious web pages found on the Internet.
URLNet processes a URL using a character-level Convolutional Neural Network (CNN)
and a word-level CNN. For the character-level CNN, the URL is first tokenized by each of the characters.

Inspired by the Xception deep object recognition model for images, Texception~\cite{Tajaddodianfar2020}
also uses separate character-level and word-level CNNs like URLNet. However, Texception's CNN kernels form different size text windows
in both the character and word levels. Multiple Texception blocks and Adaptive Max Pooling layers can be combined in
different model configurations in terms of both depth and width. In addition,
Texception utilizes contextual word embeddings in the form of either FastText or Word2Vec to convert the URL into
the input embedding vector.

Another CNN-based phishing detection model was proposed by Yerima and Alzaylaee~\cite{yerima2020high}. Using the page's content, the authors create a 31-dimensional feature vector for
each web page in their dataset and train a CNN based on this feature vector. \Sys differs from this work because it only processes the URL instead of extracting the page content which
will be much slower for inference.
Other work has proposed using LSTMs (i.e., recurrent sequential models) for phishing and malicious URL detection including~\cite{8997947,peng2019joint}. Processing LSTMs is expensive in terms of computation and memory
for long URLs which makes them impractical for large-scale production. In~\cite{huang2019phishing}, Huang et al. also investigate using capsule networks for detecting phishing URLs.

\section{Conclusion}
\label{sec:conc}
We have proposed a new transformer-based system called \Sys whose
goal is to predict the label of an unknown URL as either one which
references a phishing or a benign web page.
Transformers have demonstrated state-of-the-art performance in many natural
language processing tasks, and this paper seeks to understand if these
methods can also work well in the cybersecurity domain.

In this work, we demonstrate that transformers
which are fine-tuned using the standard BERT tasks also work remarkably well
for the task of predicting phishing URLs. Instead of extracting lexical features
or using CNNs kernels which span multiple characters and words,
which are both common in previously proposed URL detection models, our
system uses the BPE tokenizers for this task. Next, transformers convert the token sequence to an embedding
vector which can then be used as input to a standard, dense linear layer.
Results indicate that \Sys is able to significantly outperform recent
baselines, particularly over a wide range of very low false positive rates.
We also demonstrate that transformers can be made robust to novel attacks under specific threat models when we adversarially
augment the training data used for training them.

\bibliographystyle{ACM-Reference-Format}
\bibliography{urltrans}

\pagebreak
\appendix
\section{Hyperparameter Settings}
\label{sec:hyper}
For reproducibility, this appendix provides the hyperparameter settings for the three variants of the proposed \Sys model as well as those for two baseline models.
Tables~\ref{tab:URLNetParams} and~\ref{tab:TexParams} list the hyperparameters for the URLNet and Texception models that we use
as baselines in our study. The hyperparameter settings for the best performing \Sysb model are provided in Table~\ref{tab:TransParamsBert}.
In addition, the best hyperparameter settings for the \Sysr and \Sysc are given in Tables~\ref{tab:TransParamsRoberta} and~\ref{tab:TransParamsCustVoc},
respectively.

\begin{table}
\centering
\footnotesize
\begin{tabular}{|c|c|}
\hline
Parameter & Value \\
\hline
\hline
max\_len\_words & 200 \\
\hline
max\_len\_chars & 1000 \\
\hline
max\_len\_subwords & 20 \\
\hline
min\_word\_freq & 1 \\
\hline
dev\_pct & 0.001 \\
\hline
delimit\_mode & 1 \\
\hline
emb\_dim & 32 \\
\hline
filter\_sizes & {[}3,4,5,6{]} \\
\hline
default\_emb\_mode & char + wordCNN \\
\hline
nb\_epochs & 5 \\
\hline
train\_batch\_size & 128 \\
\hline
train\_l2\_reg\_lambda & 0.0 \\
\hline
train\_lr & 0.001 \\
\hline
\end{tabular}
\caption{Hyperparameters used for URLNet.}
\label{tab:URLNetParams}
\end{table}

\begin{table}
\centering
\footnotesize
\begin{tabular}{|c|c|c|}
\hline
\multicolumn{2}{|c|}{Parameter} & Value\tabularnewline
\hline
\hline
\multirow{5}{*}{
\vtop{\hbox{\strut Characters}\hbox{\strut Branch}}
} & embedding dimension & 32\tabularnewline
\cline{2-3} \cline{3-3}
 & number of blocks & 1\tabularnewline
\cline{2-3} \cline{3-3}
 & block filters & {[}2,3,4,5{]}\tabularnewline
\cline{2-3} \cline{3-3}
 & Adaptive MaxPool output & 32,32\tabularnewline
\cline{2-3} \cline{3-3}
 & maximum characters & 1000\tabularnewline
\hline
\multirow{5}{*}{
 \vtop{\hbox{\strut Words}\hbox{\strut Branch}}
} & embedding dimension & 32\tabularnewline
\cline{2-3} \cline{3-3}
 & number of blocks & 1\tabularnewline
\cline{2-3} \cline{3-3}
 & block filters & {[}1,3,5{]}\tabularnewline
\cline{2-3} \cline{3-3}
 & Adaptive MaxPool output & 32,16\tabularnewline
\cline{2-3} \cline{3-3}
 & maximum words & 50\tabularnewline
\hline
\multirow{6}{*}{
\vtop{\hbox{\strut FastText}\hbox{\strut Model}}
} & minimum words to include & 50\tabularnewline
\cline{2-3} \cline{3-3}
 & vocabulary size & 120000\tabularnewline
\cline{2-3} \cline{3-3}
 & window size & 7\tabularnewline
\cline{2-3} \cline{3-3}
 & n-grams & 2-6\tabularnewline
\cline{2-3} \cline{3-3}
 & embedding dimension & 32\tabularnewline
\cline{2-3} \cline{3-3}
 & epochs trained & 30\tabularnewline
\hline
\end{tabular}
\caption{Hyperparameters used for Texception.}
\label{tab:TexParams}
\end{table}

\begin{table}
\centering
\footnotesize
\begin{tabular}{|c|c|}
\hline
Parameter & Value \\
\hline
\hline
  attention probs dropout prob & 0.1 \\
  \hline
  hidden act & gelu \\
  \hline
  hidden dropout prob & 0.1 \\
  \hline
  hidden size & 768 \\
  \hline
  initializer range & 0.02 \\
  \hline
  intermediate size & 3072 \\
  \hline
  layer norm eps & 1e-12 \\
  \hline
  max position embeddings & 512 \\
  \hline
  num attention heads & 12 \\
  \hline
  num hidden layers & 12 \\
  \hline
  type vocab size & 2 \\
  \hline
  vocab size & 30522 \\
  \hline
  bert model & bert-base-uncased \\
  \hline
  max seq length & 128 \\
  \hline
  train batch size & 32 \\
  \hline
  learning rate & 2e-5 \\
  \hline
  num train epochs & 10 \\
  \hline
\end{tabular}
\caption{Hyperparameters  used for training the proposed Huggingface-based \Sysb model.}
\label{tab:TransParamsBert}
\end{table}

\begin{table}
\centering
\footnotesize
\begin{tabular}{|c|c|}
\hline
Parameter & Value \\
\hline
\hline
Number of Layers & 12 \\
\hline
Hidden size & 768 \\
\hline
FFN inner hidden size & 3072 \\
\hline
Attention heads & 12 \\
\hline
Attention head size & 64 \\
\hline
Dropout & 0.1 \\
\hline
Attention Dropout & 0.1 \\
\hline
Warmup Steps & 508 \\
\hline
Peak Learning Rate & 1e-4 \\
\hline
Batch Size & 2k \\
\hline
Max Epochs & 10 \\
\hline
Learning Rate Decay & Linear \\
\hline
Adam $\epsilon$ & 1e-6 \\
\hline
Adam $\beta_1$ & 0.9 \\
\hline
Adam $\beta_2$ & 0.98 \\
\hline
Gradient Clipping & 0.0 \\
\hline
Tokens per sample & 256 \\
  \hline
\end{tabular}
\caption{Hyperparameters used for fine-tuning the proposed Fairseq-based \Sysr model.}
\label{tab:TransParamsRoberta}
\end{table}

\begin{table}
\begin{subtable}{0.45\linewidth}
    \footnotesize
    \centering
    \begin{tabular}{|c|c|}
        \hline
        Parameter & Value \\
        \hline
        \hline
        Number of Layers & 3 \\
        \hline
        Hidden size & 768 \\
        \hline
        FFN inner hidden size & 3072 \\
        \hline
        Attention heads & 12 \\
        \hline
        Attention head size & 64 \\
        \hline
        Dropout & 0.1 \\
        \hline
        Attention Dropout & 0.1 \\
        \hline
        Tokens per sample & 128 \\
        \hline
        Peak Learning Rate & 1e-4 \\
        \hline
        Batch Size & 2k \\
        \hline
        Tokenizer Type & Byte BPE\\
        \hline
        Weight Decay & 0.01 \\
        \hline
        Max Epochs & 30 \\
        \hline
        \multirow{2}{*}{Learning Rate Decay} & reduce \\
         & on plateau \\
        \hline
        LR Shrink & 0.5 \\
        \hline
        Adam $\epsilon$ & 1e-6 \\
        \hline
        Adam $\beta_1$ & 0.9 \\
        \hline
        Adam $\beta_2$ & 0.98 \\
        \hline
        Gradient Clipping & 0.0 \\
        \hline
        Learning Rate & 1e-4 \\
        \hline
        vocab size & 10000 \\
      \hline
    \end{tabular}
\end{subtable} %
\begin{subtable}{0.45\linewidth}
    \footnotesize
    \centering
	\begin{tabular}{|c|c|}
        \hline
		Parameter & Value \\
		\hline
		\hline
		Learning Rate & 1e-4 \\
		\hline
		Batch Size & 2k \\
		\hline
		Max Epochs & 10 \\
		\hline
		Learning  & \multirow{2}{*}{Linear} \\
		Rate Decay & \\
		\hline
		Warmup ratio & 0.06 \\
		\hline
	\end{tabular}
\end{subtable}
\caption{Hyperparameters used for pre-training (left) and fine-tuning (right) the proposed \Sysc model.}
\label{tab:TransParamsCustVoc}
\end{table}

\end{document}